\RequirePackage{fix-cm}
\RequirePackage{amsmath}
\documentclass{svjour3}                     
\smartqed  
\usepackage[utf8]{inputenc}   
\usepackage{graphicx}
\usepackage{amsmath}
\usepackage{amssymb}
\usepackage{verbatim}      
\usepackage{geometry}
\usepackage{tikz}
\usepackage{subcaption}
\usepackage{hyperref}
\usepackage{csquotes}
\usepackage{cite}
 \journalname{Journal of Statistical Physics}
\def\no{\nonumber}
\newcommand{\tc}{\textcolor}
\begin{document}

\title{One Dimensional Exclusion Process with Dynein Inspired Hops: Simulation and Mean Field Analysis \thanks{We would like to thank Uwe C. Täuber for valuable feedback and discussions. Research was sponsored by the U.S. Army Research Office and was accomplished under Grant Number W911NF-17-1-0156. 
The views and conclusions contained in this document are those of the authors and should not be interpreted as representing the official policies, either expressed or implied, of the Army Research Office or the U.S. Government. 
The U.S. Government is authorized to reproduce and distribute reprints for Government purposes notwithstanding any copyright notation herein.}}

\author{Riya Nandi \and Priyanka}
\institute{Riya Nandi \at Department of Physics and Center for Soft Matter and Biological Physics, Virginia Tech, Blacksburg, VA 24061-0435, USA. \email{riya11@vt.edu}, \and Priyanka \at  Department of Physics and Center for Soft Matter and Biological Physics, Virginia Tech, Blacksburg, VA 24061-0435, USA. \email{pri2oct@vt.edu}}

\date{Received: date / Accepted: date}

\maketitle

\begin{abstract}
We introduce a one-dimensional non-equilibrium lattice gas model representing the processive motion of dynein molecular motors over the microtubule. We study both dynamical and stationary state properties for the model consisting of hardcore particles hopping on the lattice with variable step sizes. We find that the stationary state gap-distribution exhibits striking peaks around gap sizes that are multiples of the maximum step size, for both open and periodic boundary conditions. We verified this feature using a mean-field calculation. For open boundary conditions, we observe intriguing damped oscillator-like distribution of particles over the lattice with a periodicity equal to the maximum step size. To characterize transient dynamics, we measure the mean square displacement that shows weak superdiffusive growth with exponent $\gamma \approx 1.34$ for periodic boundary and ballistic growth ($\gamma \approx 2$) for open boundary conditions at early times. We also study the effect of Langmuir dynamics on the density profile.


\end{abstract}
\maketitle

\section{Introduction}
\label{intro}
Biological molecular motors are protein molecules that transport molecular cargos within the living cells by moving progressively along actin filaments and microtubules \cite{Schliwa:2004, Howard:2001}. The term \enquote{motor} is motivated by the fact that molecular motors utilize the chemical energy produced by the hydrolysis of ATP (Adenosine triphosphate) to ADP(Adenosine diphosphate) to perform mechanical work. This progressive movement of motors is often considered as analogous to the traffic in a city, where the cytoskeletal filaments act as the path along which the molecular motors travel in a directed fashion \cite{kolomeisky2015motor}.  The efficient transport of molecular motors is critical for the healthy functioning of a cell as they play crucial roles in many biological processes such as cell division, transfer of genetic information, etc. Therefore, understanding the motor dynamics constitutes an important and relevant research area. Among the three known motor families, kinesins and dyneins move along a microtubule, whereas myosins move on an actin filament~\cite{Schliwa:2003n}. Unlike the other two, cytoplasmic dynein mediates retrograde transport shuttling organelles, vesicles, etc. from the edge to the body of the cell. Owing to several experiments and theoretical modeling of the dynamics of motors in the last few years, we have a reasonably good understanding of the mechanism of processive backward/forward motion of kinesin and myosin. However, the functioning of cytoplasmic dynein is much less understood. Advancement in probing techniques in recent years has shed some light on the structural complexity and motility of dynein motors. Moreover, experimental studies of dynein have shown that they exhibit an unusual gear-like mechanism taking variable step sizes of $8$, $16$, $24$, and $32$ nm depending on the concentration of ATPs available and external loads ~\cite{BHABHA:2016, mallik2004cytoplasmic}. 

Motivated by this jump strategy with variable step-size, in this paper we introduce a simple one-dimensional stochastic model, with the goal to understand the effect of this motility mechanism on the collective dynein dynamics. Both analytical and numerical stochastic modeling of statistical physics is predominantly used to understand wide varieties of biological phenomena ranging from transport to cell division~\cite{fisher1999force, chou2011non, julicher1997modeling, wang2002ratchets}. For such studies, particle-based modeling \cite{Spitzer:1970} constitutes a simplified, idealized but powerful approach where one ignores the structural complexity of individual biological motors and considers them as hardcore particles that jump (move) across lattice sites (linear microtubule tracks). The process of hydrolysis is a constant source of energy to the motors, which can be represented in the model as a constant driving force for the particles. Such a driving force prevents the system from ever relaxing to thermal equilibrium. Thus, after an initial transient period of non-equilibrium dynamics, the system asymptotically evolves to a non-equilibrium stationary state \cite{fisher1999force}. In 1968, a very simple one-dimensional model, the totally asymmetric simple exclusion process (TASEP), was first used to model the collective movement of ribosomes on messenger RNA track \cite{MacDonald:1968}. Later TASEP variants were introduced to study molecular motors~\cite{chou2011non}; however, very few of the earlier studies are based on dynein motors~\cite{Kunwar:2006, Mukherji:2008, singh2005monte,Ashwin:2010}. An earlier study used a stochastic cellular automata model to represent the collective behavior of dynein motors~\cite{Kunwar:2006}. However, this study did not incorporate the effect of load and hindrance together. To understand the dependence of the collective behavior of dynein motors on both load and blockage due to traffic, we formulate a TASEP variant. We specifically ask how dyneins optimize step sizes depending on parameters such as load and the presence of other motors~\cite{singh2005monte, rai2013molecular}. In this approximate model of dynein transport, which incorporates the essential load-dependent hopping mechanism, we investigate their long-time as well as the transient time behavior. Although open boundary conditions represent a more realistic model for a system of biological motors, we decide to study the system with periodic boundary conditions first since the bulk properties of the system are not affected by this choice. Later we look at the boundary effects by simulating our model with open boundary conditions. We find that the stationary-state gap distribution in front of the motors exhibits intriguing peaks at multiples of the maximum jump size, with the amplitude of these peaks decaying exponentially as a function of the gap size. For open boundary conditions, we observe a damped oscillatory density profile starting from the entry sites. Depending on the influx and escape rates and the rates of attachment/detachment, these oscillations decay as one moves further into the bulk leading to a flat density profile. We also analyze the superdiffusive growth of the mean square displacement during the transient time followed by standard diffusive behavior.

Further, some experimental observations have revealed that under the effect of forces in the direction opposing the motion of the motors, they unbind from the tracks, thus decreasing their processivity~\cite{visscher1999single,coppin1997load}. Inspired by the phenomena of \enquote{finite processivity}, we perform simulations for the open boundary conditions with Langmuir kinetics~\cite{parmeggiani2004totally}. In-vivo studies of dyneins, in particular, show that for low load, the probability of unbinding increases with load. However, at higher load, the dyneins exhibit catch-bond type behavior~\cite{kunwar2011mechanical} which results in the probability of unbinding decreasing with increasing load. Here we want to emphasize that our goal is not to mimic the exact functional form of the load-dependence but to probe the effect of adding a non-zero probability of evaporation and deposition of particles on the steady-state distributions of our exclusion process as a mere exercise.

This paper is structured as follows: In the following section, we describe our stochastic model and review the known gear-like jump mechanism of dynein motors. We then discuss the results for the static and dynamics of the model for periodic boundary conditions in Section \ref{PBC}. Section \ref{OBC} is devoted to the discussion of the properties of the model for open boundary conditions. Following that, we discuss the effects of adding Langmuir kinetics on the open boundary steady-state properties in Section \ref{evdep}. We conclude with a summary of our results and open questions in Sec. \ref{Conc}.
\begin{figure}
    \centering
    \includegraphics[width=0.5\columnwidth]{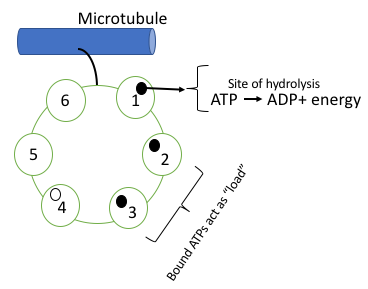}
    \caption{Schematic of a single head of dynein attached to the microtubule. Each head has six ATP binding sites (AAA1-6).}
    \label{Figure:1}
\end{figure}

\section{Model and Simulation Method}
\label{sec:1}
In this section, we briefly review the structure and the mechanism of motion of dynein motors; we refer the reader to Ref.~\cite{BHABHA:2016} for a detailed description. Cytoplasmic dyneins perform unidirectional discrete jumps on the microtubule. A single microtubule consists of approximately $13$ protofilaments, and it has been assumed that dynein uses more than one protofilament to walk due to its large structure \cite{DeWitt:2012,Qiu:2012}.
Structurally, a dynein motor has two heads; each head contains 6 AAA (ATPases Associated with diverse cellular Activities) domains, four of which are potentially ATP binding sites, as shown in Fig.~\ref{Figure:1}. Earlier studies of dynein motors show that the hydrolysis (conversion of ATP to ADP) occurs on all ATP binding sites; however, only the hydrolysis at the primary site AAA1 is compulsory for the movement of the motors, so the motors do not hop until the primary site hydrolyzes. The attached ATP units on the secondary sites act as an external load for the motors, which determine the step size over the microtubule and also influence other molecular functions~\cite{mallik2004cytoplasmic, singh2005monte}. In our study, we simplified the system by considering that dyneins with stacked head walk on a single protofilament, which has been also observed previously in certain in-vitro experiments~\cite{Shibata:2012} where they demonstrate that multiple rows of parallel protofilaments are not essential for the motility of the dyneins. In their Monte-Carlo studies, Mallik et al.~\cite{mallik2004cytoplasmic}, have also previously argued that considering the dynein as a single-headed structure is a reasonable assumption. 

In our study, we introduce a variant of the totally asymmetric exclusion process to model the dynamics of dynein motors. We consider $L$ discrete lattice sites with $N$ hardcore particles which we refer to as $N$-motor system. A lattice site can be occupied by at most one particle, and each particle jumps in the forward direction with varying step sizes of $1,~2,~3$ or $4$ depending on the number of ATP units attached to the secondary sites (loads) of the particle. Each particle carries a flag for the occupancy of one primary and three secondary ATP binding sites. Additionally, we also assume that there is an infinite source of ATP units available for binding. We employ random sequential Monte-Carlo update where, at each time step of the simulation, a motor is chosen randomly from an occupied lattice site. The hopping probability with varying step size is determined by three distinct processes:\\
\textbf{ATP attachment-} One unit of ATP attaches to the empty primary site with a constant probability $P_{on}$ or to any of the available secondary sites, with a constant probability $S_{on}$. \\
\textbf{ATP detachment-} One unit of ATP detaches either from the primary or one of the three secondary sites with constant probabilities $P_{off}$ and $S_{off}$, respectively. \\
\textbf{ATP hydrolysis-} A filled primary binding site hydrolyzes and converts ATP to ADP. This chemical reaction leads to energy production, which propels the motor forward. The number of steps the motor moves depends on the number of ATP units bound to the secondary sites and the number of available empty lattice sites in front of the motor. If the secondary ATP binding sites are unoccupied, the motor attempts to hop to the maximum step size of four. If $s$ number of secondary sites hold ATP ($s$ can be 1, 2, or 3), given the primary binding site is occupied, the motor attempts to hop $(4-s)$ steps. The motor can only take a step forward as long as there is no other motor blocking its path. If $n$ is the attempted hopping size and $m$ is the number of empty lattice sites in front of the motor, it takes $\min(m,n)$ steps.
In actual biological motors, even the secondary sites hydrolyze ATP to ADP, but this does not result in forward-propulsion of the motor. In our $N$-motor system, secondary site hydrolysis is not considered. However, one can argue that the detachment rate of AAA(2-4) takes care of the hydrolysis of ATP from secondary sites.   

The stochastic probabilities of attachment and detachment $P_{on},~S_{on},~P_{off},~S_{off}$ are in general considered to be functions of the stall force, the temperature of the cellular environment and concentration of available ATP\cite{mallik2004cytoplasmic}. However, for this study, we choose constant probabilities. Based on some of the in-vitro experimental observations~\cite{reck2004molecular,kon2004distinct} where the attachment rate of ATP to the primary site is large in comparison to the ATP attachment rate to secondary sites, we will present most of our results for high attachment rates to AAA1, specifically for the case when, $S_{on}=1-P_{on}$,~$P_{off}=1-P_{on}$,~$S_{off}=P_{on}$. All simulation results are obtained for lattice size $L=10^3$ for periodic boundary conditions and $L=500$ for open boundary conditions. Ensemble averaging of $10^4$ different realizations have been performed for all final results.

For the maximal step size $m_s$, the rate of hopping $n$ steps with $m$ empty sites in front of a particle is given by, 
\begin{eqnarray}
  u_n(m)&=&\sum_{i=1}^{m_s-1}\delta_{n,i}[v_{2i-1}\delta_{m,i}+v_{2i}\theta(m-(i+1))]+v_{2m_s-1}\theta(m-m_s)\delta_{n,m_s}~,
\label{rate}
\end{eqnarray}
where {the rates $v_j$ are the constant rate of hopping which depend on the number of ATP molecules attached to the binding sites. $v_{2n}$ denote the rate of jumping $n$ steps if $m>n$ empty sites are in front of the particle and $v_{2n-1}$ is the rate of jumping $n$ steps if exactly $n$ empty sites is in front of the particle. Also $\theta(x)$ is a discontinuous function whose value is zero for $x < 0$ and one for $x \ge 0$. The terms within the summation give the hopping rates for jump lengths smaller than $m_s$, and the last term denotes the hopping rate for maximum jump length, $m_s$.} 

\begin{figure*}
    \begin{subfigure}[c]{0.5\textwidth}
     \centering
     \includegraphics[width=\columnwidth]{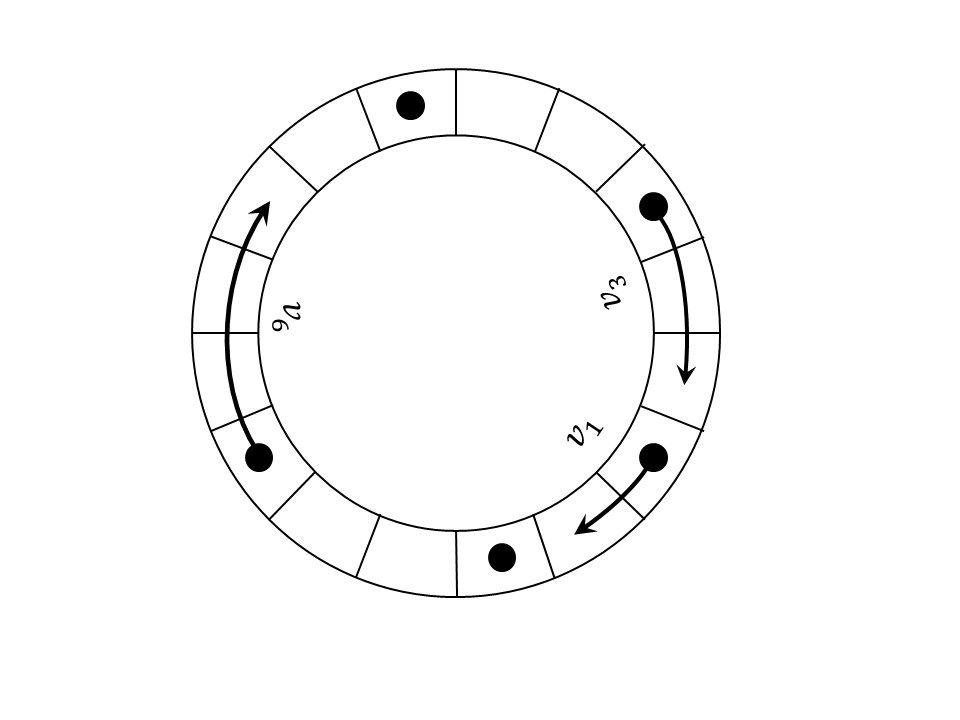}
    \end{subfigure}
    \hspace{-1cm}
   \begin{subfigure}[c]{0.5\textwidth}
    \centering
    \includegraphics[width=\columnwidth]{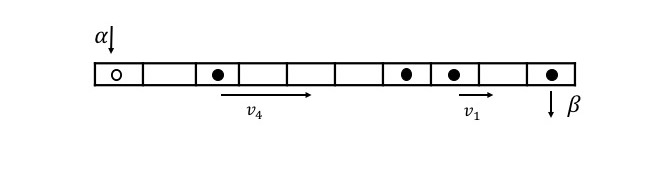}
   \end{subfigure}
   \caption{Transport of dynein motors on a lattice with periodic boundary conditions (left) and with open boundary conditions (right). For open boundary conditions, motors enter and exit the lattice with influx and escape rates $\alpha$ and $\beta$, respectively. Motors hop forward (clockwise) $n$ steps with probability $v_n$ and $m$ ($m>n$) empty sites in front of them.}
   \label{Figure:2}
\end{figure*}

{In our model of dynein motors, where a motor can take maximum four steps ($m_s=4$),}
\begin{eqnarray}
  u_n(m)&=&[v_1\delta_{m,1}+v_2\theta(m-2)]\delta_{n,1}+[v_3\delta_{m,2}+v_4\theta(m-3)]\delta_{n,2}\no\\&&+[v_5\delta_{m,3}+v_6\theta(m-4)]\delta_{n,3}+v_7\delta_{n,4}\theta(m-4)~,
\label{rate4}
\end{eqnarray}
{For $m_s=4$, approximated functional form $v_i$ in term of the attachment and detachment rate of ATPs' to the binding sites are given as,
\begin{eqnarray}
 v_7&\approx& P_{on}(1-P_{off})\left[\sum_{i=0}^{3}(S_{on}S_{off})^{i}(1-S_{on})^{3-i}\right]~,\\
 v_6&\approx&  P_{on}(1-P_{off})S_{on}(1-S_{off})[(S_{on}S_{off})^{2}+(1-S_{on})^2+(1-S_{on})S_{on}S_{off}]~,\\
 v_5&\approx& v_7\rho+v_6~,\\
 v_4&\approx&  P_{on}(1-P_{off})S_{on}^2(1-S_{off})^{2}[(1-S_{on})+S_{on}S_{off}]~,\\
 v_3&\approx& (v_7+v_6)\rho+v_4~,\\
 v_2&\approx& P_{on}(1-P_{off})[S_{on}^3(1-S_{off})^3]~,\\
 v_1&\approx&(v_7+v_6+v_4)\rho+v_2~.
\end{eqnarray}
 We have also written down the exact form of all the rates for $m_s=2$ in the Appendix equation (\ref{rate_rel}).}
Here, we aim to understand the effect of various rates of attachment/detachment, and of the motor density on the fluctuations and stationary state properties of the system. 
We have studied the system for both periodic and open boundary conditions shown in Fig.~\ref{Figure:2}. For open boundary conditions, the same dynamical rules of hopping apply in the bulk as for the periodic boundary case. Moreover, the open ends of the lattice are attached to a reservoir through which a motor can enter and exit with the rates $\alpha$ and $\beta$ from the first and last site, respectively.

\section{Results}
\subsection{Periodic Boundary Conditions}
\label{PBC}
\subsubsection{Steady-state properties:}
\begin{figure}[t]
\begin{subfigure}[t]{.48\textwidth}
 \centering 
   \includegraphics[width=\columnwidth]{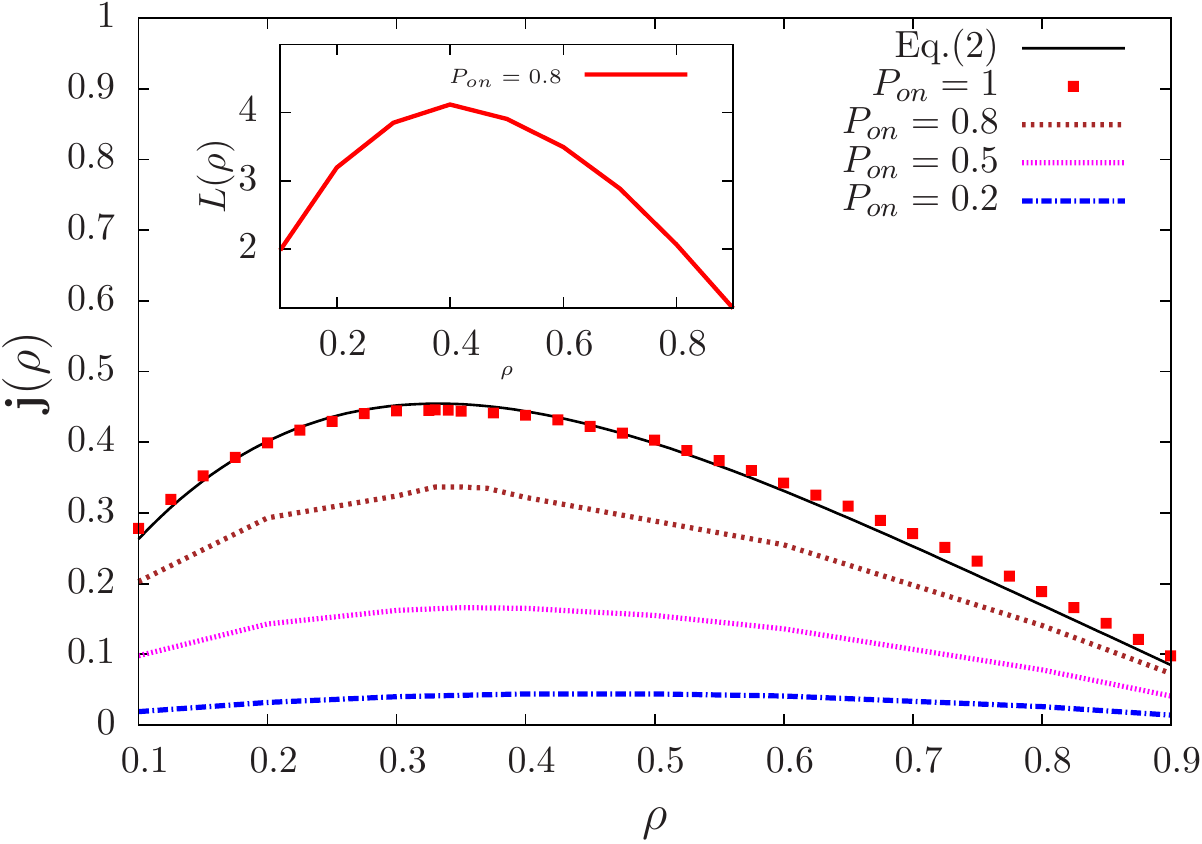}
   \caption{}
 \end{subfigure}
 \hfill
 \begin{subfigure}[t]{0.48\textwidth}
 \centering 
   \includegraphics[width=\columnwidth]{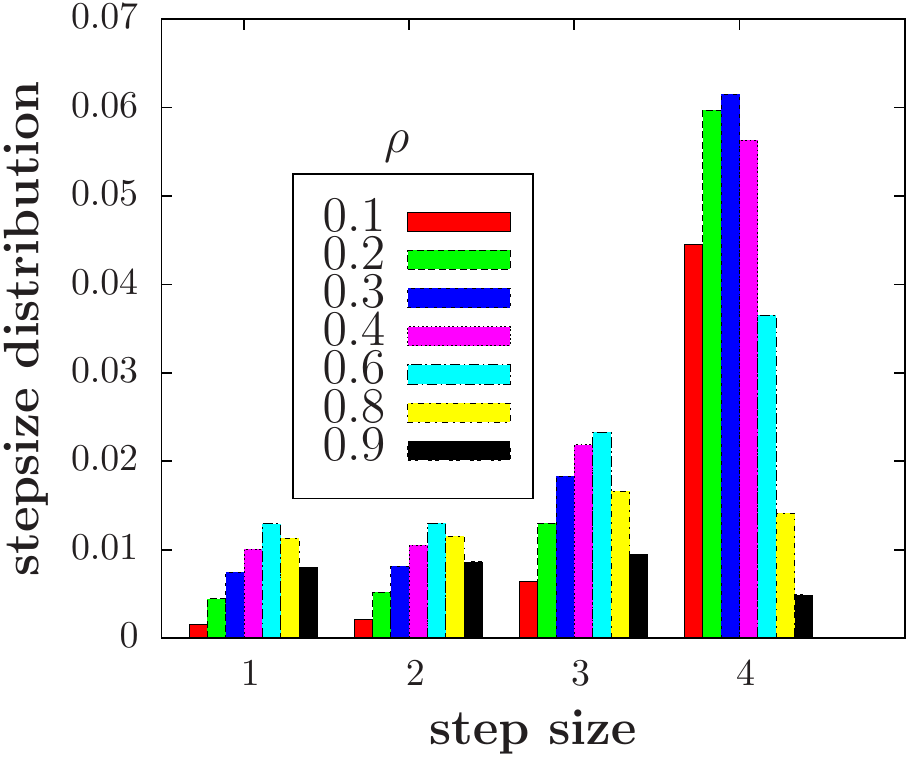}
    \caption{}
 \end{subfigure}
 \caption{(a) Current profile $j(\rho)$ as a function of density for four different rates of attachment and detachment. The numerically obtained data points (red points) are compared with the mean-field Eq.~\ref{current} (black line) for $P_{on}=1$, the maximum current is obtained at $\rho = 0.33$. The inset shows the flux due to \enquote{load} (ATP units carried by the secondary sites) $L(\rho)$, reaches maxima at density $\rho\sim0.4$ for $P_{on}=0.8$. (b) The distribution of the hop sizes performed by the motors in the stationary state peaks at hop size four for all densities. Attachment and detachment rates used in the simulation are $P_{on}=~S_{off}=~0.8$, $P_{off}=~S_{on}=~0.2$.} 
 \label{Figure:3}
\end{figure}

The system reaches it's stationary state when the particle current does not change with time, which implies $\frac{d j}{dt}=0$. we numerically calculate the time scale after which the particle current as well as the density profile become constant. In our model the steady-state is achieved by evolving the system for approximately $L^2$ Monte-Carlo steps. 
{The stationary state current across the bond $i-1$ and $i$ is the sum of all the possible jumps and is given by
\begin{eqnarray}
j(\rho)&=&\langle v_1\eta_{i-1}(1-\eta_{i})\eta_{i+1}+v_3\eta_{i-1}(1-\eta_{i})(1-\eta_{i+1})\eta_{i+2}+...+\no\\
&&v_2 \eta_{i-1}(1-\eta_{i})(1-\eta_{i+1})\sum_{l=0}^{\infty}\eta_{i+l+2}\prod_{k=1}^{l}(1-\eta_{i+1+k})+...\rangle~,
\end{eqnarray}}
where $\eta_j$ is occupancy of site $j$. We have also solved for the stationary state properties of the system in the mean-field approximation where we ignored all higher-order correlations, and for the two-point correlation, considered the factorization {$\langle\eta_i\eta_j\rangle=\langle \eta_i\rangle\langle \eta_j\rangle=\rho^2$. 
This approximation manifests in two contexts the current work; one is to calculate the particle current and the other to calculate the mean-field gap distribution later in the paper where we consider the empty spaces are uncorrelated.} One can define the mean-field particle current as
\begin{equation}
j(\rho)=\rho\sum_{i=1}^{m_s}\Big[ a_i\sum_{l=1}^i(1-\rho)^l\Big]~,
\label{current}
\end{equation}
 where, {$ a_i$ is the probability of a particle to jump $i$ steps and its value ranges between $0-1$. For $m_s=4$,$a_1=v_1+v_2,~a_2=v_3+v_4,~a_3=v_5+v_6$ and $a_4=v_7$.} Further for the limiting case of $P_{on}=S_{off}=1,~S_{on}=P_{off}=0$, where the particles can take only four steps. The above expression for the current then simplifies to,
\begin{equation}
 j(\rho)=\tc{red}{v_7}\rho\sum_{i=1}^4(1-\rho)^i~.
\end{equation}
We then calculate the maxima of the current by setting $d j(\rho)/d\rho=0$. For the limiting case, when {$v_7=1$}, the maximum current is obtained at density $\rho_{max}=0.3312$. This is verified by our simulation result where the current profile for $P_{on}=1$ indeed peaks at a $\rho \approx 0.33$ as can be seen from Fig.~\ref{Figure:3}(a). We also observe that for all our chosen rates, the current reaches peak value at densities below $\rho=0.5$. This shift of the peak below half-density signifies the absence of particle-hole symmetry. Simply put, in this system a particle moving four steps from site $i$ to $i+4$, is not equivalent to a hole moving from site $i+4$ to $i$. This shift of maximum density below half also suggests the self-assembly of particles and holes to optimize the jump efficiency. To validate this claim, we investigate the flux due to \enquote{load} or ATP carried by the secondary sites, $L(\rho)$. For $P_{on}=0.8$, $L(\rho)$ attains maximum value at $\rho\sim 0.4$ which is only slightly different from the density where density peak for the same choice of the rates, see inset of Fig.~\ref{Figure:3}(a). Setting $S_{on}$ equal to zero, and $m_s$ to one, we recover the well-known TASEP results~\cite{Spitzer:1970}. 

\begin{figure}[b]
\centering
 \begin{subfigure}[t]{.48\textwidth}
 \centering
  \includegraphics[width=\columnwidth]{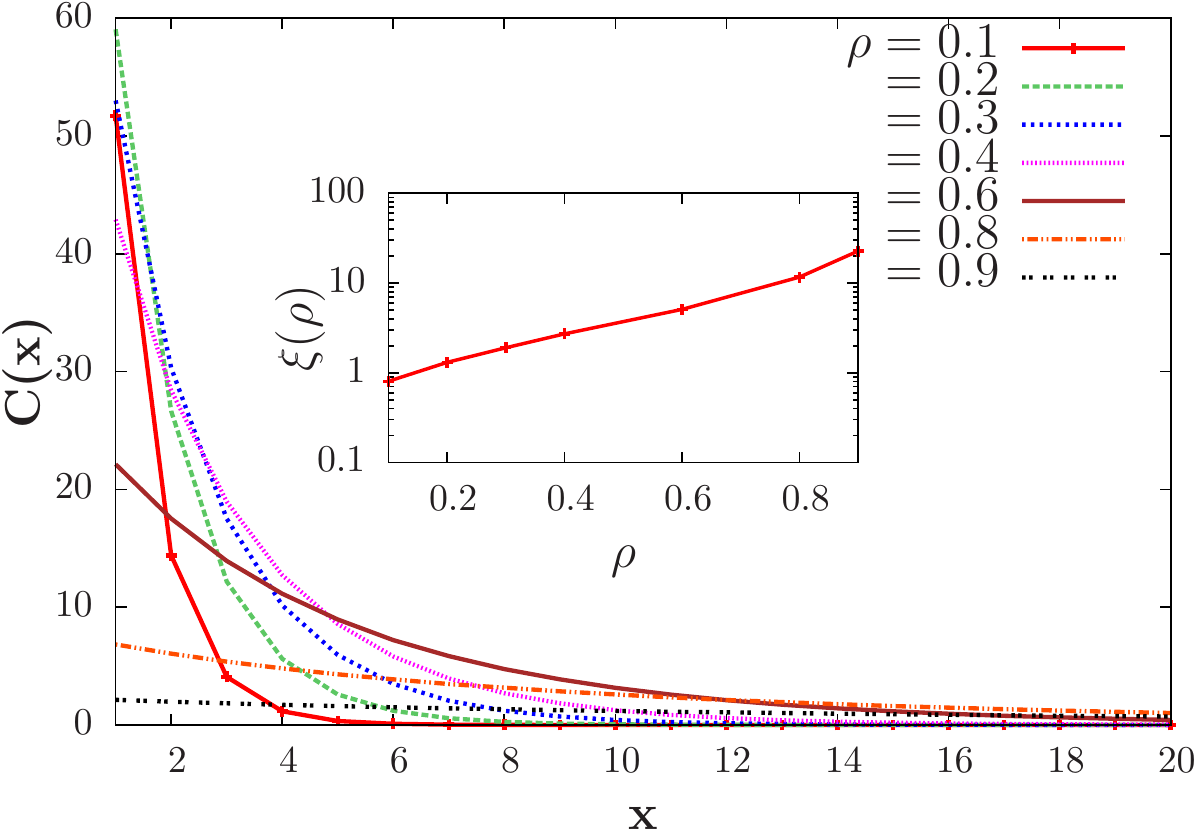}
    \caption{}
\end{subfigure}%
\hfill
\begin{subfigure}[t]{0.48\textwidth}
 \centering
  \includegraphics[width=\columnwidth]{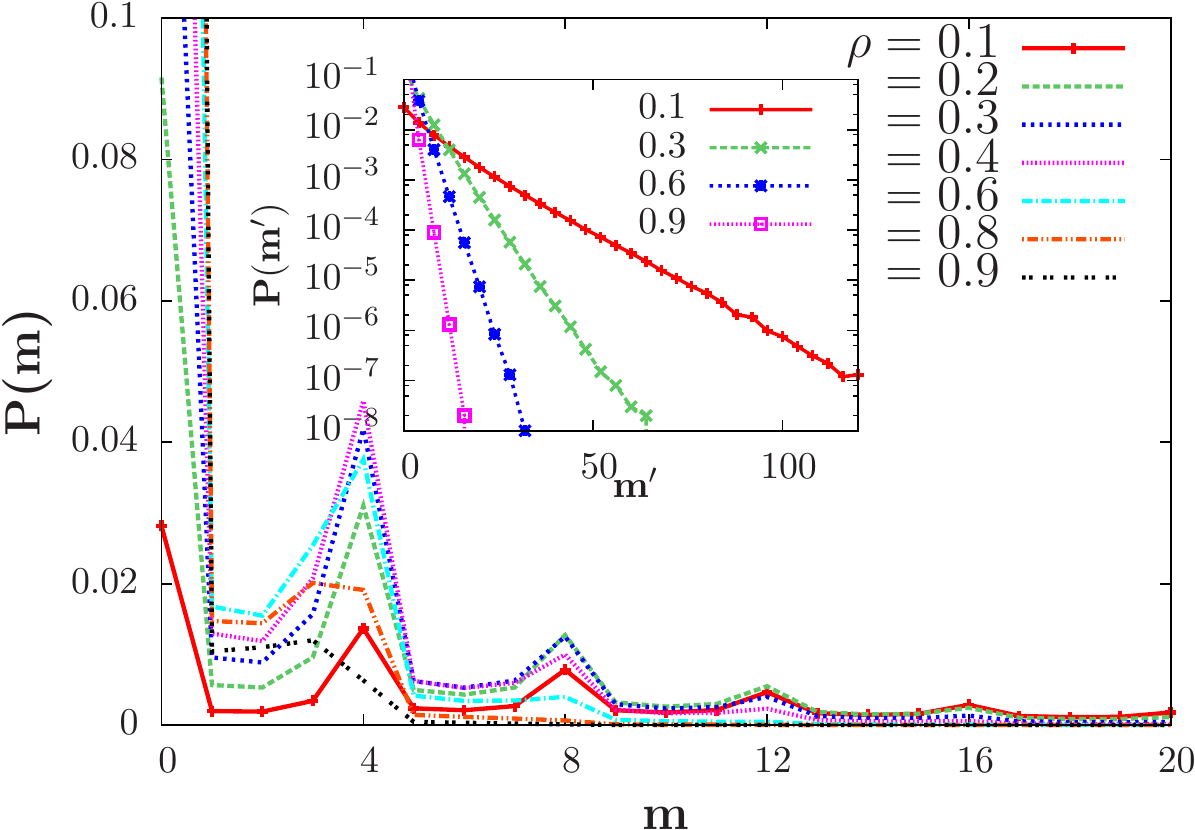}
    \caption{}
 \end{subfigure}
\caption{(a)Particle cluster distribution $C(x)$ as a function of cluster size x for different densities {show exponetial decay}. Inset shows exponential growth of correlation length as a function of density. (b) The stationary-state gap distribution $P(m)$ shows peaks at multiples of four gap sizes for various densities. Inset shows the exponential decay of the distribution of the peaks $P(m\prime)$ for all densities. The rates of attachment and detachment here are $P_{on}=~S_{off}=~0.8$, $P_{off}=~S_{on}=~0.2$.}
\label{Figure:4}
\end{figure}

In order to further investigate this self-organization of the particles, we probe into the distribution of particle clusters as well as the distribution of empty spaces in front of the particles referred to as {\it{gap distribution}}, $P(m)$. In the stationary state, the particle cluster {is an exponentially decaying function} for all choices of rates at all densities, as shown in Fig.~\ref{Figure:4}(a). The characteristic size of the cluster depends not only on the density but also on the chosen rates of attachment and detachment. One can also extract the correlation length from the slopes for the log-linear plot of cluster distribution. The calculated correlation length is a measure of the size of the particles' clusters that are distributed over the entire lattice and we find that these clusters grow exponentially with density owing to the absence of long-range correlations. Moreover, the gap distribution shows a very striking feature for our specific choice of $P_{on},~S_{on},~P_{off}$, and $S_{off}$: The distribution peaks at multiples of the maximum possible step size $m_s$, and the amplitude of the peaks decay exponentially, for all densities as can be seen in the inset of Fig.~\ref{Figure:4}(b). The exponential decay of the distribution of the gaps with density is just complementary to the exponential growth of the correlation length. We inspected this further by simulating the gap distribution for various $m_s$, and the characteristic features of the peaks observed are consistent. These peaks signify that the motors assemble themselves in such a way that they optimize the utilization of energy by covering the maximum possible steps during each jump. The amplitude of these peaks at exact gap sizes which are multiples of $m_s$ is reduced with an increase in $S_{on}$ which increases load and consequently the probability of the motors jumping by smaller step sizes. Thus, these peaks can be attributed to the smaller attachment rate of secondary sites which favors the large step-size of $m_s$. This is also verified in figure~\ref{Plotm_s2} for various choices of probabilities. This observation is supported by the step-size distribution plot \ref{Figure:3}(b) as well, where one can see the particles indeed prefer to hop by the maximum length of four steps rather than taking shorter steps of one, two or three for all range of densities. We also notice that the step size distribution of four steps peaks at the density $\rho \approx 0.3$. 

In order to analytically calculate the gap distribution, 
{we assume that the system has  mean-field geometry where all the motors are connected to each other. In this limit, all the higher order correlations are ignored and the joint gap distribution ${\cal P}(m_1, m_2, m_3,...m_n)$ of $n$ particles having $m_1,m_2,..m_n$ holes in front of them can be factorized i.e., ${\cal P}(m_1, m_2,...,m_n) =\prod_i^{n} P(m_i)$}. We write down the evolution equation for the probability of finding a gap of size $m$ for a totally biased walk in the mean-field approximation as~\cite{Kunwar:2006, Godreche:2016, Shrivastav:2010}:
\begin{align}
    \frac{dP(m,t)}{dt}=&-\Big[\sum_{k=1}^m u_k(m)+\sum_{j=1}^\infty P(j,t)\sum_{k=1}^j u_k(j)\Big]P(m,t) +\sum_{k=1}^{m_s} P(m+k,t)u_{k}(m+k)\no\\
  & +\sum_{k=1}^m P(m-k,t)\sum_{j=k}^\infty P(j,t)u_k(j)~ \qquad (m \ge 1)~, \label{modes_c} \\
    \frac{dP(0,t)}{dt} =& -P(0,t)\sum_{j=1}^\infty P(j,t)\sum_{k=1}^ju_k(j)+ \sum_{j=1}^{m_s} P(j,t)\sum_{k=1}^j u_{k}(j)~.
\label{MF}
\end{align}
The above equations must satisfy the following conservation rules, 
\begin{align}
\sum_{m\ge 0}P(m,t)&=1~, \label{Crule1}\\
\sum_{m\ge 1}mP(m,t)&=\frac{1}{\rho}-1~,
\label{Crule2}
\end{align}
where $\rho=N/L$ is the density of motors present in the system. The conservation rules imply that we have a closed system with no global particle number fluctuations. To handle the infinite set of equations (\ref{MF}), we rewrite it into a single differential equation using the generating function technique where we multiply both sides of Eq.~(\ref{MF}) by $z^m$, sum over all $m\ge1$, and define the generating function $G(z, t)=\displaystyle \sum_{m\ge1}z^{m}P(m,t)$. For the case of $m_s=4$, the further calculation becomes cumbersome. Hence we carry out the analysis for a smaller step size $m_s=2$ and arrive at the simplified form of the master equation in the stationary state. The detailed calculation and comparison of results with Monte Carlo simulations for maximum step size $m_s=2$ are presented in \ref{appA}. For any arbitrary jump size, the steady-state gap distribution has a polynomial solution and takes the following form for rate Eq.~\ref{rate}~\cite{Shrivastav:2010}
\begin{equation}
P(m)=\sum_{i=1}^{m_s}\alpha_i q_i^{m}~.
\end{equation}
We can also define the current in terms of the gap distribution as 
\begin{equation}
j(\rho)=\rho\sum_{k=1}^{m_s}P(k)\sum_{n=1}^{k}u_n(k)~.
\end{equation}
In addition to the static steady-state quantities, we also study the stationary state dynamical behavior by determining the mean square displacement of a tagged particle in this modified TASEP-like model. Earlier studies have shown that the variants of TASEP in one dimension fall in KPZ universality class, where dynamic exponent $z=3/2$~\cite{Shamik:2007, Priyanka:2016b}. We calculate the stationary state mean square displacement of tagged particles defined as $\langle\delta \sigma^2(t)\rangle =\langle d^2\rangle-\langle d\rangle^2~,$ where $d$ is the average displacement traveled by the tagged particle and the angular bracket is an average over the different realizations starting from the same initial conditions for this N-motors system. Our simulation results show the same characteristic features as observed in the case of 1D TASEP~\cite{Shamik:2007}. We see the initial KPZ growth with an exponent $2/3$ which is a signature of the TASEP universality class, followed by diffusion. The standard finite-size scaling form shows perfect scaling collapse with dynamical exponent $z_s=3/2$, as shown in Fig.~\ref{Figure:5}(a).
\subsubsection{Transient behavior:}
Next, we investigate the dynamics of the system starting from random initial configurations and calculate the mean square displacement of the motors in time,
\begin{equation}
\langle \delta r^2(t)\rangle =\langle r^2\rangle-\langle r\rangle^2~,
\end{equation}
where $\displaystyle r(t)=\frac{1}{N}\sum_{i=1}^N x(t)$ is the average displacement of the motors over the entire lattice and the angular bracket represents ensemble averaging.
\begin{figure}[t]
    \centering
    \begin{subfigure}[t]{0.48\textwidth}
    \centering
     \includegraphics[width=\columnwidth]{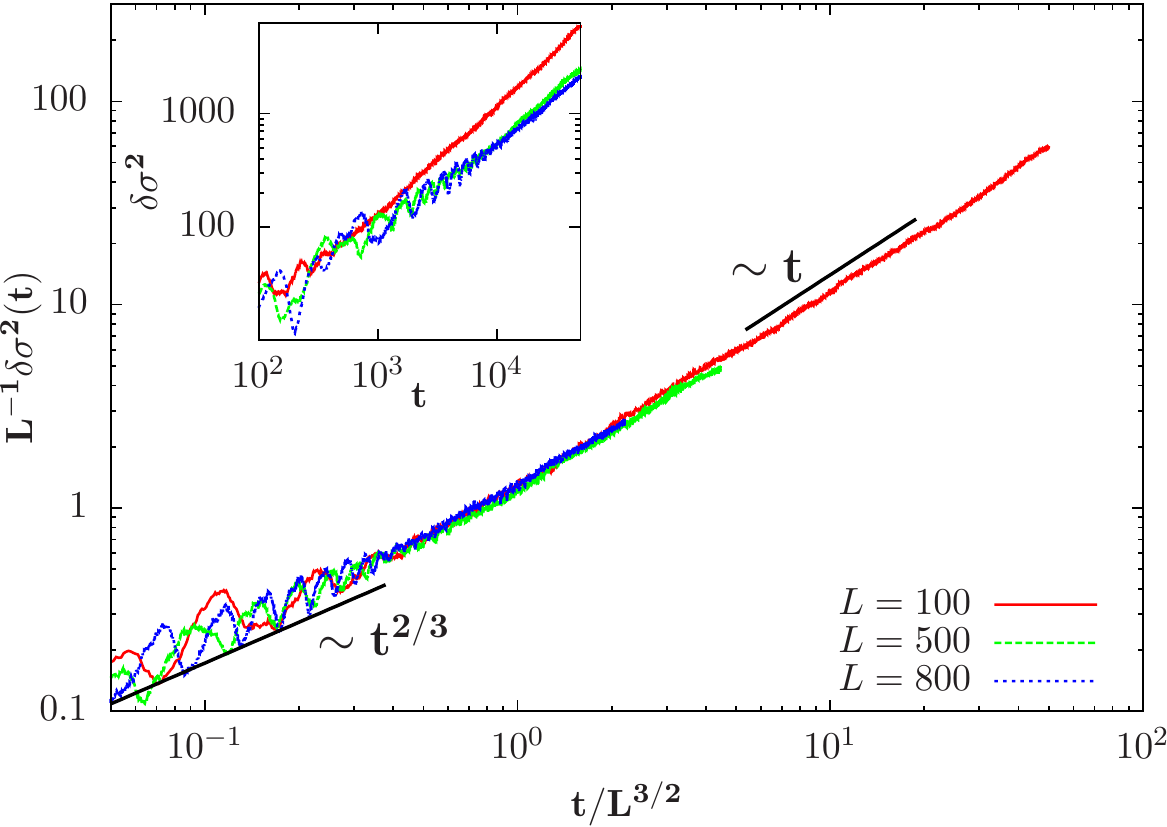}
    \end{subfigure}
    \hfill
    \begin{subfigure}[t]{0.48\textwidth}
     \centering
     \includegraphics[width=\columnwidth]{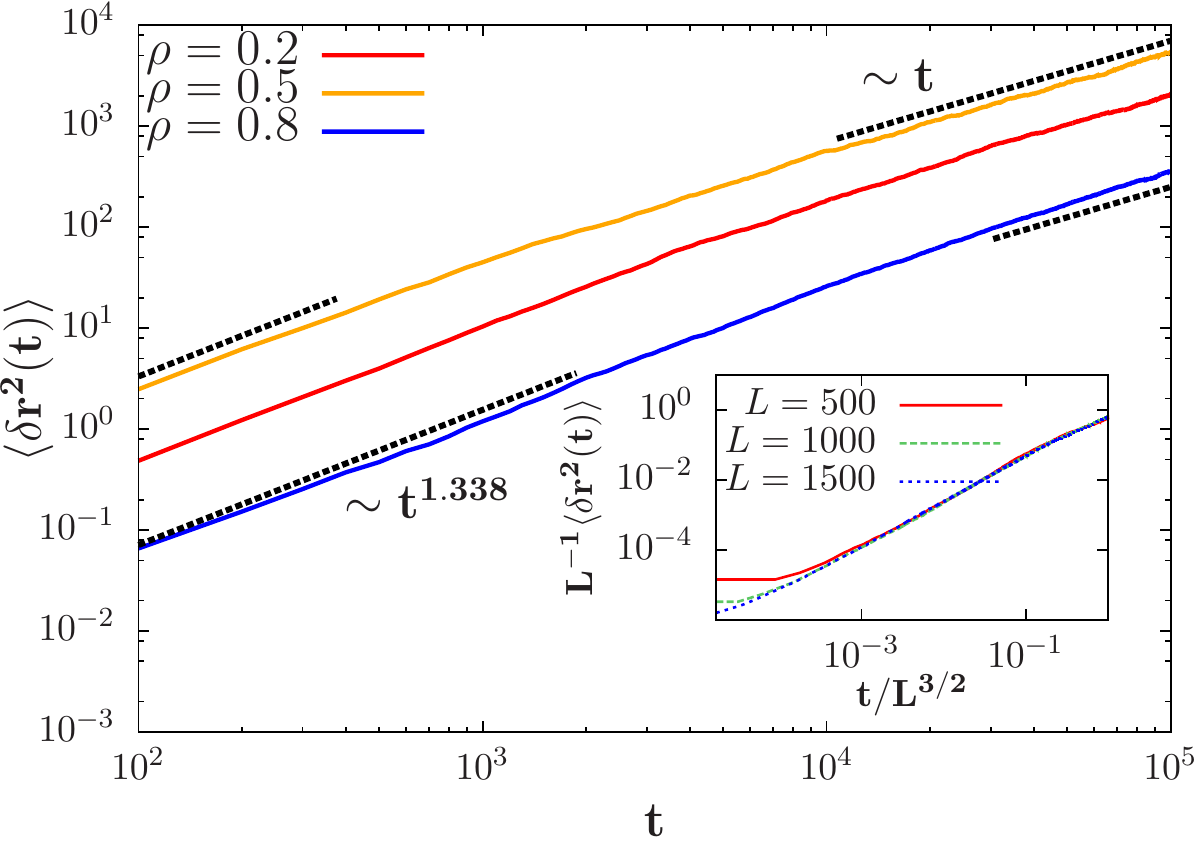}
    \end{subfigure}
    \caption{(a) Growth of the mean square displacement for a tagged particle with time in the stationary state. (b) Superdiffusive growth of the mean square displacement with time starting from a random initial condition at early times. Inset shows system size scaling with exponent $z=3/2$. Rates of attachment and detachment chosen here are, $P_{on}=S_{off}=0.8$, $P_{off}=S_{on}=0.2$.}
    \label{Figure:5}
\end{figure}

We measure the mean square displacement for various densities as well as for different attachment and detachment rates. They all display power-law behavior with time $ \langle\delta r^2(t) \rangle \sim t^{\gamma}$ as shown in  Fig.~\ref{Figure:5}(b). After a small initial time window $t_{mic}\sim 100$ time steps, the intermediate time growth of mean square displacement shows anomalous diffusion, which means faster than diffusive behavior. The best fit of the transient regime gives growth exponent $\gamma \approx 1.338\pm 0.01$ before it crosses over to a standard diffusive regime at later times where $\gamma$ is one. We also observe that the crossover time depends on the density of the motors present in the system. This kind of super-diffusive behavior has been observed experimentally in the dynamical regime of molecular motors and other active cellular matter~\cite{bruno2009transition,lau2003microrheology,bursac2005cytoskeletal}. In most biological systems, short-range fluctuations are responsible for the diffusive propagation, whereas long-range fluctuations arising from directed transport driven by chemical energy cause super-diffusive dynamics\cite{sharma2019unusual,caspi2000enhanced}. One can interpret the crossover from super-diffusive to diffusive behavior in our simulations with a similar argument. At early times, the motion is primarily dictated by individual hops resulting from ATP hydrolysis and uncorrelated collective dynamics. This weak super-diffusive power-law could perhaps also be attributed to logarithmic correction. After the transient period, the dominant stochastic collisions between the particles slow down the growth of mean square displacement, which results in standard diffusive behavior in the long-time regime. We also evaluate the dynamic scaling exponent using the finite-size scaling relation,
\begin{equation}
    \langle\delta r^2(t)\rangle \propto L^{\xi}f\Big(\frac{t}{L^z}\Big).
    \label{scaling}
\end{equation}
The best data collapse gives the exponents $\xi=1$ and $z=3/2$, as shown in the inset of Fig. \ref{Figure:5}(b)).

\subsection{Open Boundary Conditions}
\label{OBC}
 In this section, we discuss the results and analysis of the collective behavior of an N-motor system with open boundary conditions. The open boundary lattice is a more realistic depiction of the biological system as microtubules are open-ended tracks on which a motor can enter and exit from the left end (head) and the right end (tail), respectively. A motor enters with probability $\alpha$ provided that the first site is empty and exits from last lattice site with probability $\beta$. The rules of hopping in bulk are the same as in the periodic boundary case. However, if the particle reaches the last lattice site while hopping with the ATP dependent jump probability, then it exits the lattice. We perform simulations with three different rates of influx and escape: $\alpha=2\beta,~\alpha=\beta$, and $~\alpha=\beta/2$, starting with different initial densities.  For arbitrary $m_s$, one can write the per-site occupancy dynamics for density at the $k^{th}$ site $\rho_k$ as
\begin{align}
\frac{d\rho_1}{dt}=&\alpha(1-\rho_1)-v_s\rho_1(1-\rho_{2})~, \\
\frac{d\rho_k}{dt}=&\Big[\bar{v}\rho_{k-1}+\tilde{v}\rho_{k-2}(1-\rho_{k-1})\Theta(k-1)\Big](1-\rho_{k})\rho_{k+1}+\no\\
 &\sum_{j=1}^{\text{min}(k-1,m_s)}v_j\rho_{k-j}\prod_{i=0}^{j-1}(1-\rho_{k-i})-
 v_s\rho_k(1-\rho_{k+1})~,\qquad 1< k< L \\
 \frac{d\rho_L}{dt}=&\sum_{j=1}^{m_s}v_j\rho_{L-j}\prod_{i=0}^{j-1}(1-\rho_{L-i})-\beta\rho_L~,
 \end{align}
 
where $v_s=\sum_{j=1}^{m_s}v_j$, $\bar{v}=v_2+v_3+v_4$, $\tilde{v}=v_3+v_4$, and $\alpha,~\beta$ are the entry and exit rates, respectively.

\begin{figure}[t]
 \begin{subfigure}[t]{0.48\textwidth}
 \centering
   \includegraphics[width=\columnwidth]{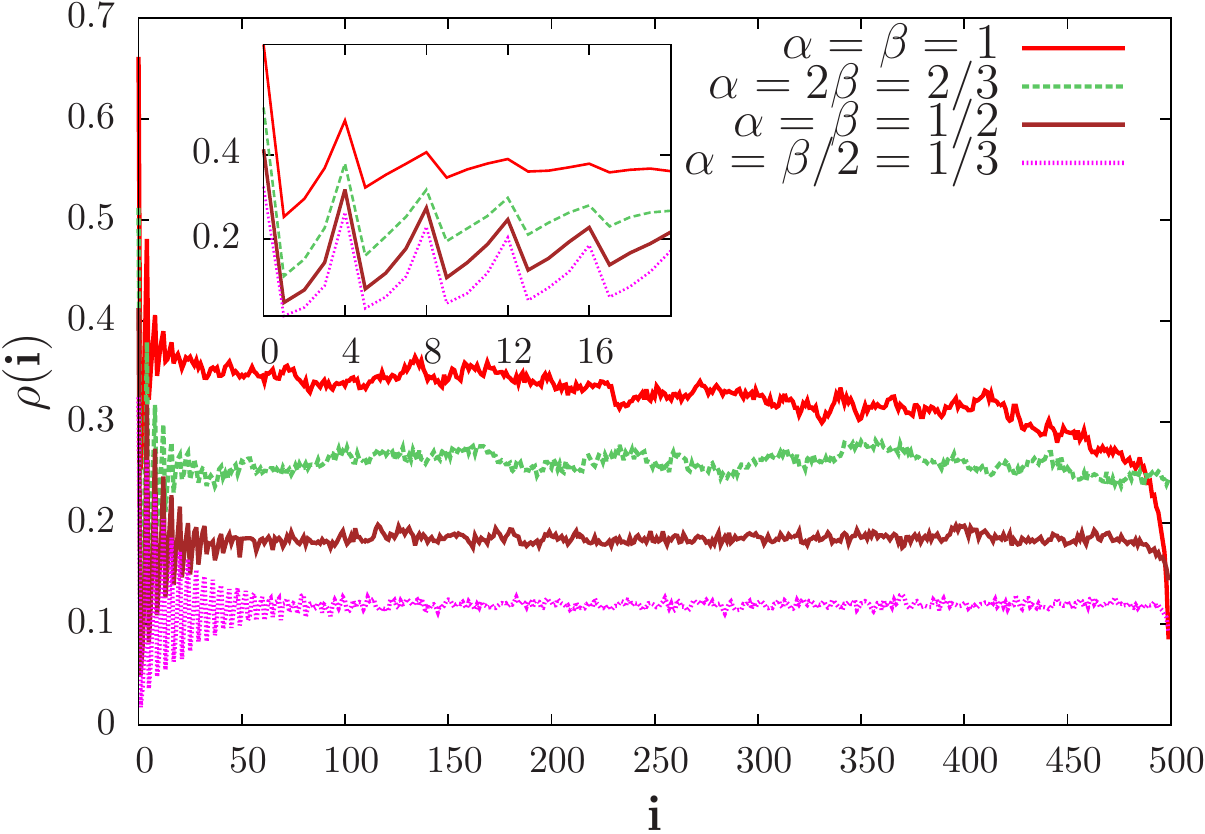}
    \caption{}
 \end{subfigure}
 \hfill
 \begin{subfigure}[t]{0.48\textwidth}
 \centering 
   \includegraphics[width=\columnwidth]{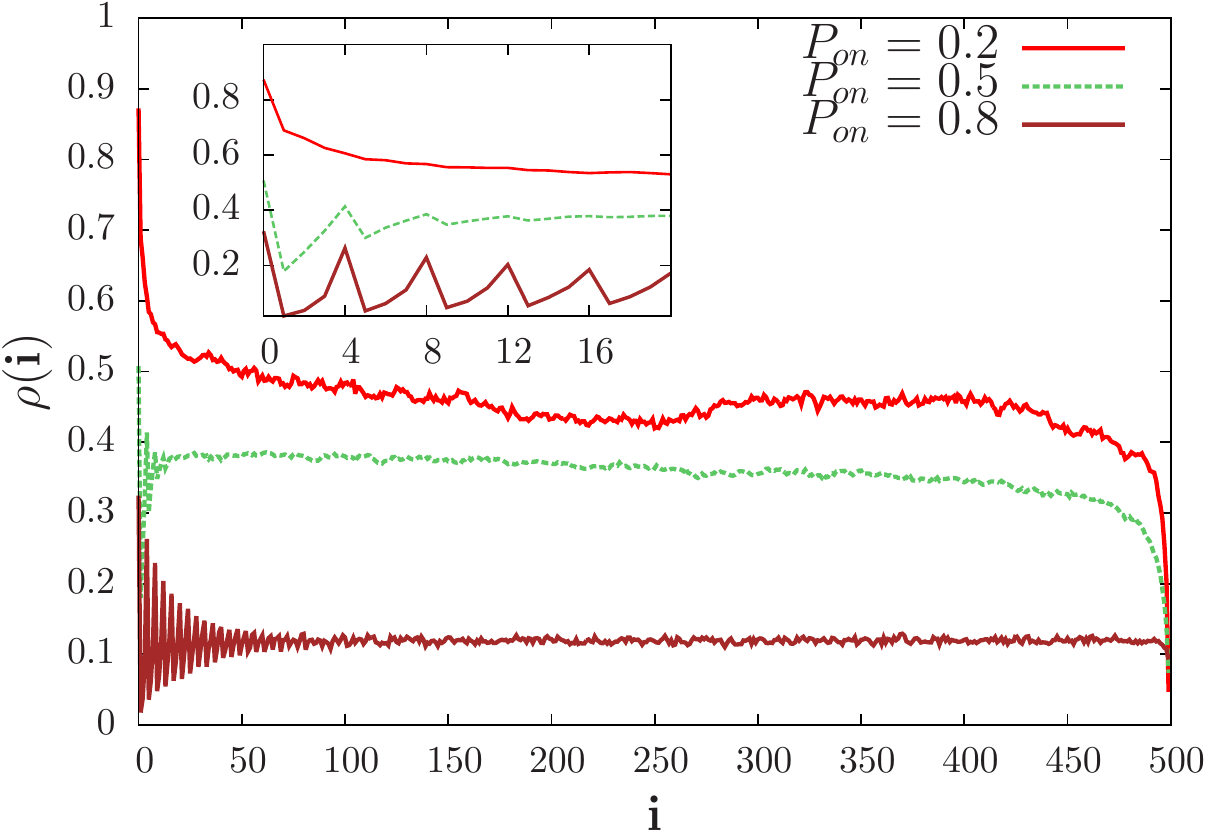}
    \caption{}
 \end{subfigure}
 \caption{(a) Stationary state density profile across the lattice sites for different influx and escape rates $\alpha$ and $\beta$, for $P_{on}=S_{off}=0.8$, $P_{off}=S_{on}=0.2$. The inset shows oscillations with periodicity of $m_s=4$ in the density profile near the boundary through which the motors enter the system. (b) The density profile for different rates of attachment/detachment for $\alpha=2\beta=2/3$ shows similar periodic behavior near the entrance boundary.}
 \label{Figure:6}
\end{figure}

\subsubsection{Steady state properties:}
We first analyze the stationary state density profile over the lattice. Interestingly, as can be seen in Fig.~\ref{Figure:6}, we observe oscillatory behavior near the boundary, where the particles enter the system. The average density peaks at those lattice sites which are multiples of the maximum jump size $m_s$, with the amplitude of the peaks decaying exponentially as one moves further into the bulk. The characteristic decay length depends on the influx rate; the larger $\alpha$, the faster the decay. The depth of these oscillations into the bulk of the system also decreases with the increasing influx rates, see \ref{Figure:6}(a). Significantly high influx rate causes crowding near the entry site forcing the motors to take smaller steps. The depth of oscillations into the bulk also decreases upon lowering $P_{on}$, as shown in Fig.~\ref{Figure:6}(b). Reducing $P_{on}$ implies an increase in the attachment rate $S_{on}$ of secondary ATP binding sites, which consequently increases the probability of taking smaller step sizes. Thus the amplitude of these oscillations decays and vanishes faster displaying a TASEP-like tangent profile~\cite{schutz2000exactly}. From our simulation results, we conclude that the mean density in bulk never exceeds $\rho=0.5$, even for significantly high influx rates. For large $P_{on}$ and small $S_{on}$ values, the bulk density profile is almost flat. The gap distribution shows similar peaks at multiples of the maximum step-size with exponentially decaying amplitudes as discussed for the case of periodic boundary conditions, see Fig.~\ref{Figure:7}(a).

\subsubsection{Transient behavior:} 
\begin{figure}[t]
 \begin{subfigure}[t]{0.48\textwidth}
 \centering
  \includegraphics[width=\columnwidth]{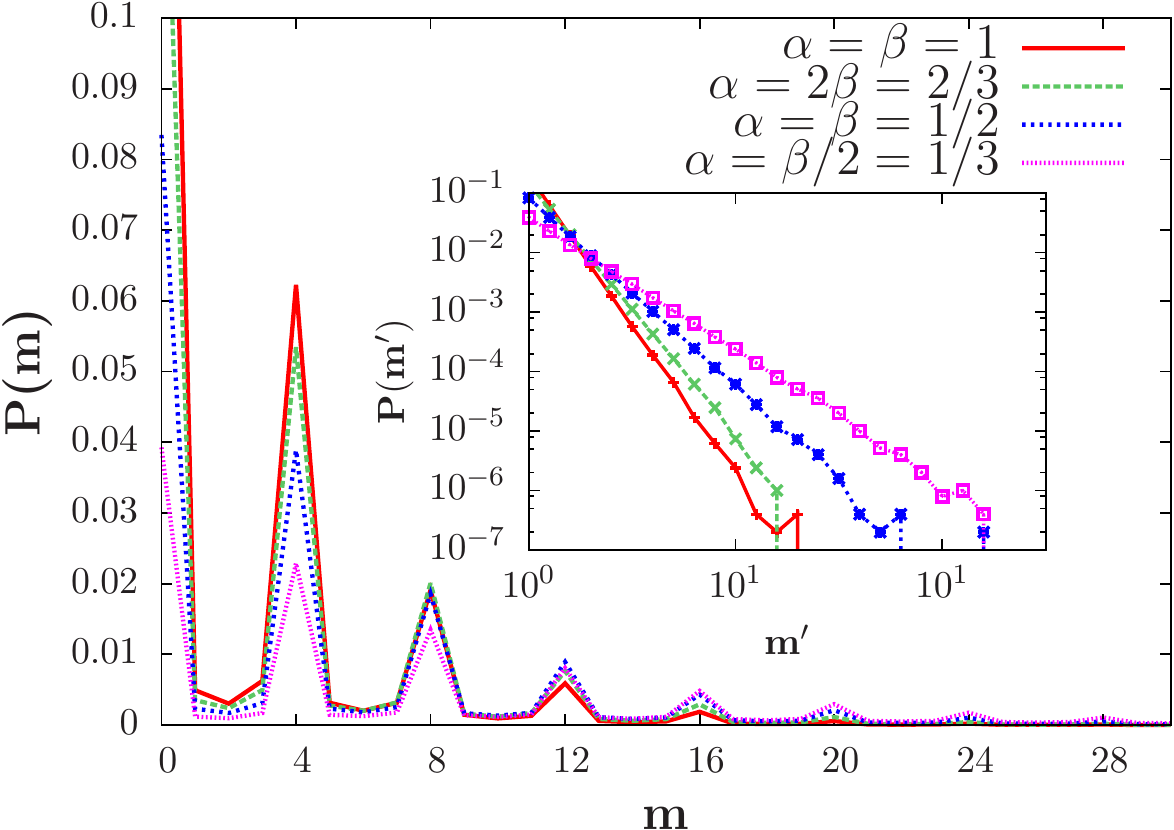}
    \caption{}
 \end{subfigure}
 \hfill
 \begin{subfigure}[t]{0.48\textwidth}
 \centering 
   \includegraphics[width=\columnwidth]{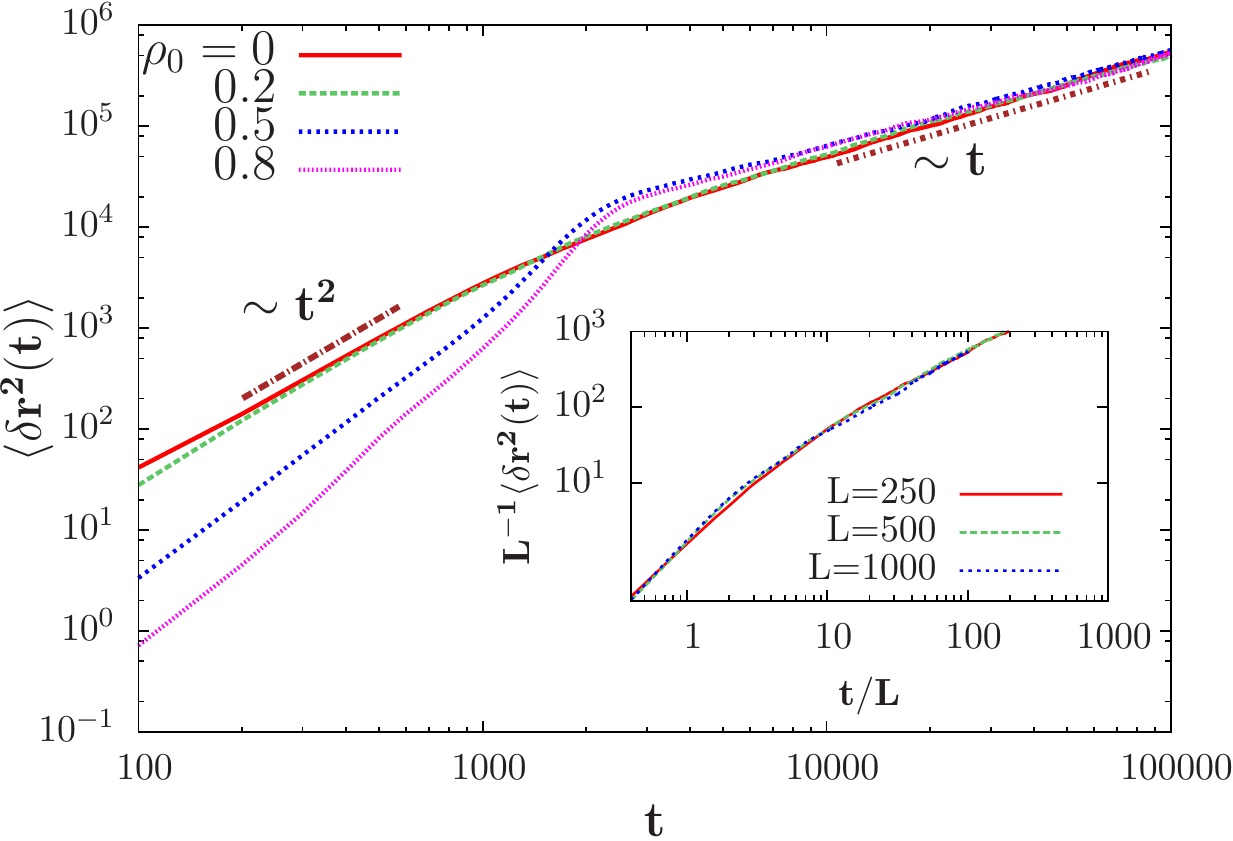}
    \caption{}
 \end{subfigure}
 \caption{(a)The stationary state gap distribution $P(m)$ shows peaks at multiples of $4$ for different rates $\alpha$ and $\beta$, and the inset shows the exponential decay of these peaks with gap size $m$. (b) Growth of fluctuations for different starting densities, for the case $\alpha=2\beta=2/3$. The inset shows the finite size scaling collapse of mean square displacement with the dynamical exponent, $z=1$.}
 \label{Figure:7}
\end{figure}

To understand the dynamics of the system, we furthermore analyze the mean square displacement $\langle\delta r^2(t)\rangle$ during the transient regime, as was done for periodic boundary conditions in sec. \ref{PBC}. During the growth of the mean square displacement, we observe three different growth regimes; initial time dynamic exhibits a combination of super-diffusive and ballistic growth ($\langle\delta r^2(t)\rangle \sim \Gamma t^{1.2}+ \Gamma' t^2$, where $\Gamma$ and $\Gamma$' are fitting parameters) followed by complete ballistic growth ($\langle\delta r^2(t)\rangle \sim t^2$) in the intermediate regime. The long-time behavior of the system is predominantly diffusive. This ballistic growth preceding the diffusive stationary state can be explained by the accelerated motion of the particles in the intermediate regime as there is enough empty space present in front of them. We also probe the effect of different starting densities on the growth of fluctuations. As shown in Fig.~\ref{Figure:7}(b), for zero initial density one observes a more prolonged ballistic regime; with the increase in initial density, the pure ballistic regime becomes negligible. For finite initial density, we observe prolonged mixed ballistic and diffusive dynamics. However, in the long-time regime, the dynamic behavior is independent of the initial densities, and the effect of stochastic forces start to become evident, forcing the motors to undergo diffusion \cite{huang2011direct,galanti2013persistent}. Furthermore, plots for different system sizes collapse according to the finite-size scaling relation~(\ref{scaling}) with exponents $\xi=1$ and $z=1$, as shown in the inset of Fig. \ref{Figure:7}(b)). The dynamical exponent $z=1$ is the resultant of ballistic growth of fluctuations.     

\subsection{Effect of Langmuir Kinetics}
\label{evdep}
Hitherto our analysis was based on the assumption that dynein motors do not unbind from the microtubule track; in other words, they exhibit \enquote{infinite processivity}. But experiments have shown that molecular motors often unbind from the tracks with a load-dependent probability~\cite{visscher1999single,coppin1997load}. Loose motors can also bind to the microtubule with a constant probability. The unbinding of motors from the filaments under load has been explored via various theoretical frameworks~\cite{parmeggiani2001detachment,kafri2005dynamics}. In the present study, we analyze the effect of unbinding on the collective dynamics of the dyneins by introducing a load-dependent probability of evaporation $P_{ev}$. Since in our model the load is represented by the number of ATP molecules attached to the secondary sites, we model the probability of evaporation to be an exponential function $P_{ev}=\frac{e^{s-3}}{L}$, where $s=(0,1,2,3)$ represents the number of ATPs bound to the secondary site~\cite{ohashi2019load} and $L$ is the lattice size, a prefactor chosen to maintain the evaporation at a low threshold. We choose the constant probability of deposition such that it is equal to the lowest evaporation probability, $P_{dep}=\frac{e^{-3}}{L}$.   

\begin{figure}[b]
 \begin{subfigure}[t]{0.48\textwidth}
 \centering
  \includegraphics[width=\columnwidth]{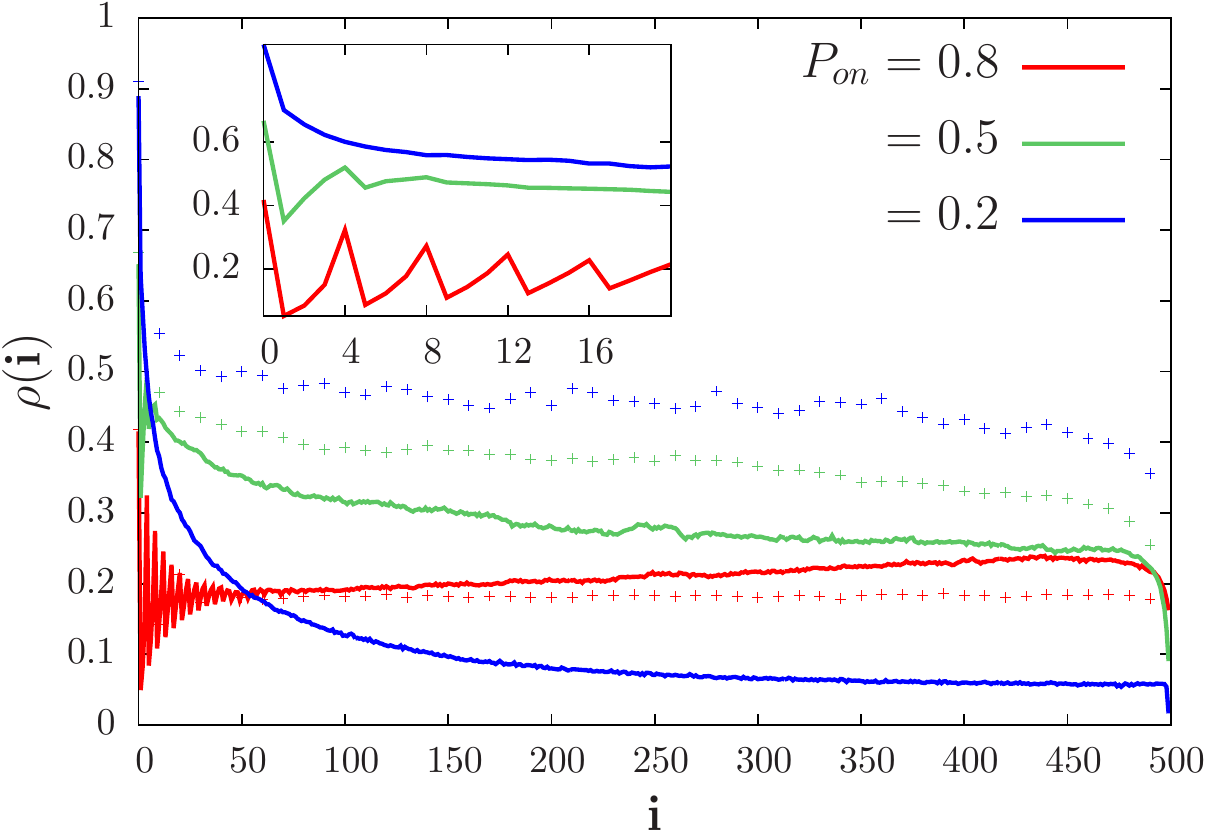}
    \caption{}
 \end{subfigure}
 \hfill
 \begin{subfigure}[t]{0.48\textwidth}
 \centering 
   \includegraphics[width=\columnwidth]{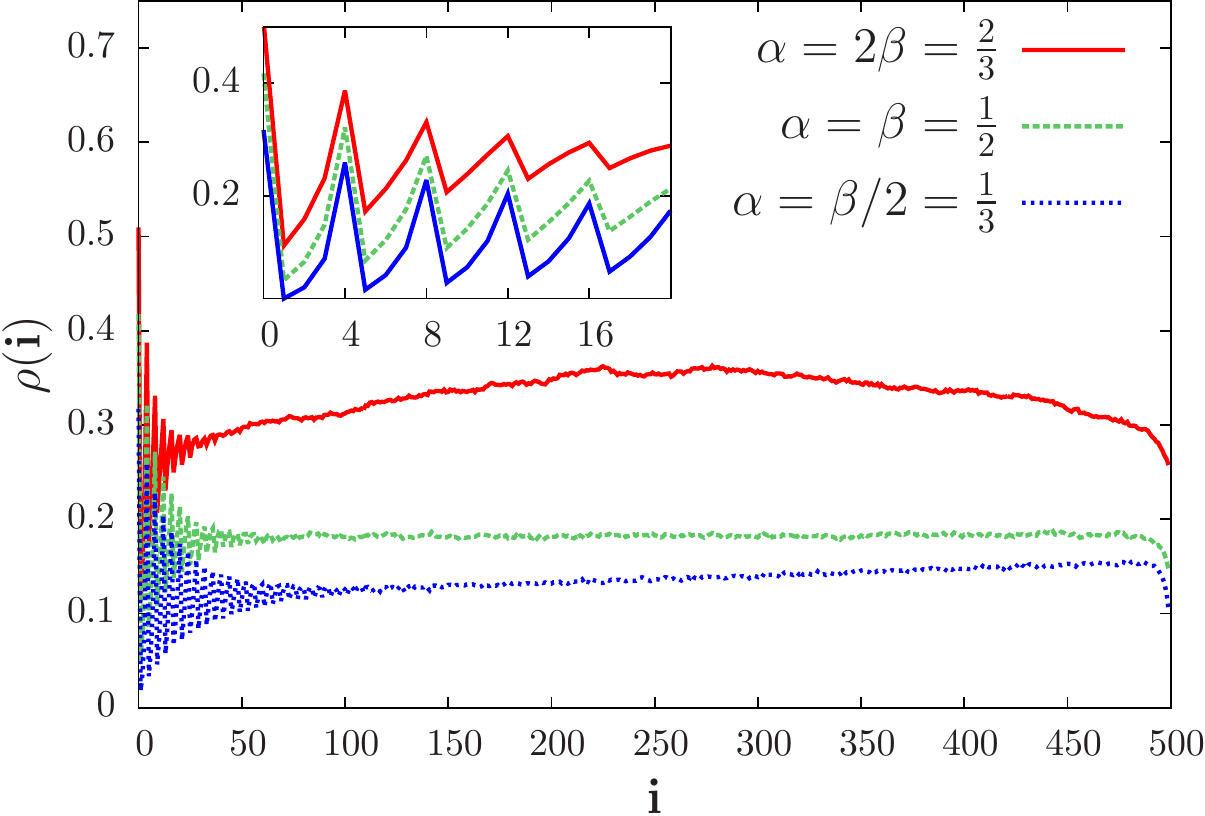}
    \caption{}
 \end{subfigure}
 \caption{(a) Open boundary stationary state density profile for different rates of attachment/detachment. The solid lines represent data with Langmuir kinetics and the points represent the corresponding data without Langmuir kinetics. The inset shows oscillations with periodicity four near the entrance boundary. The chosen probabilities of influx and escape are $\alpha=\beta=0.5$.
 (b) Density profile for different probabilities of influx and escape for the probability of attachment/detachment $P_{on}=S_{off}=0.8,~S_{on}=P_{off}=0.2$.} 
 \label{Figure:8}
\end{figure}

\begin{figure}[b]
 \begin{subfigure}[t]{0.48\textwidth}
 \centering
  \includegraphics[width=\columnwidth]{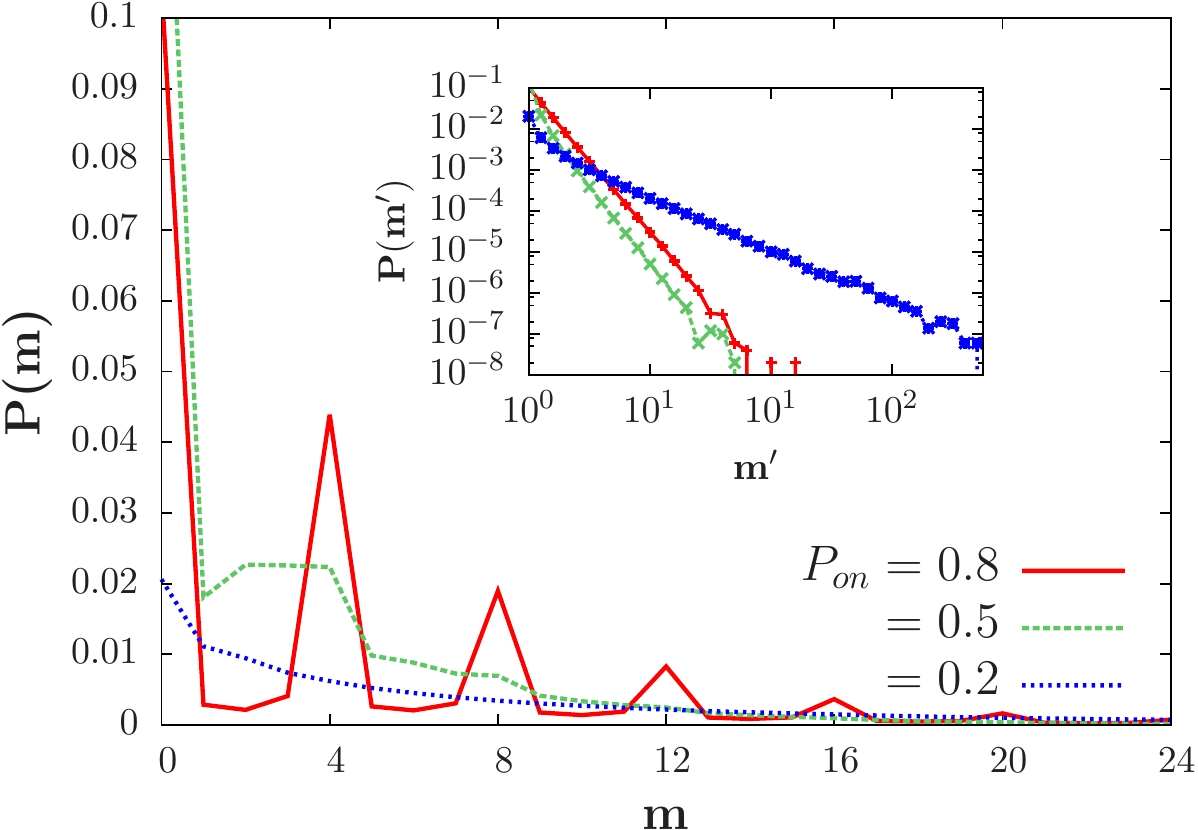}
    \caption{}
 \end{subfigure}
 \hfill
 \begin{subfigure}[t]{0.48\textwidth}
 \centering 
   \includegraphics[width=\columnwidth]{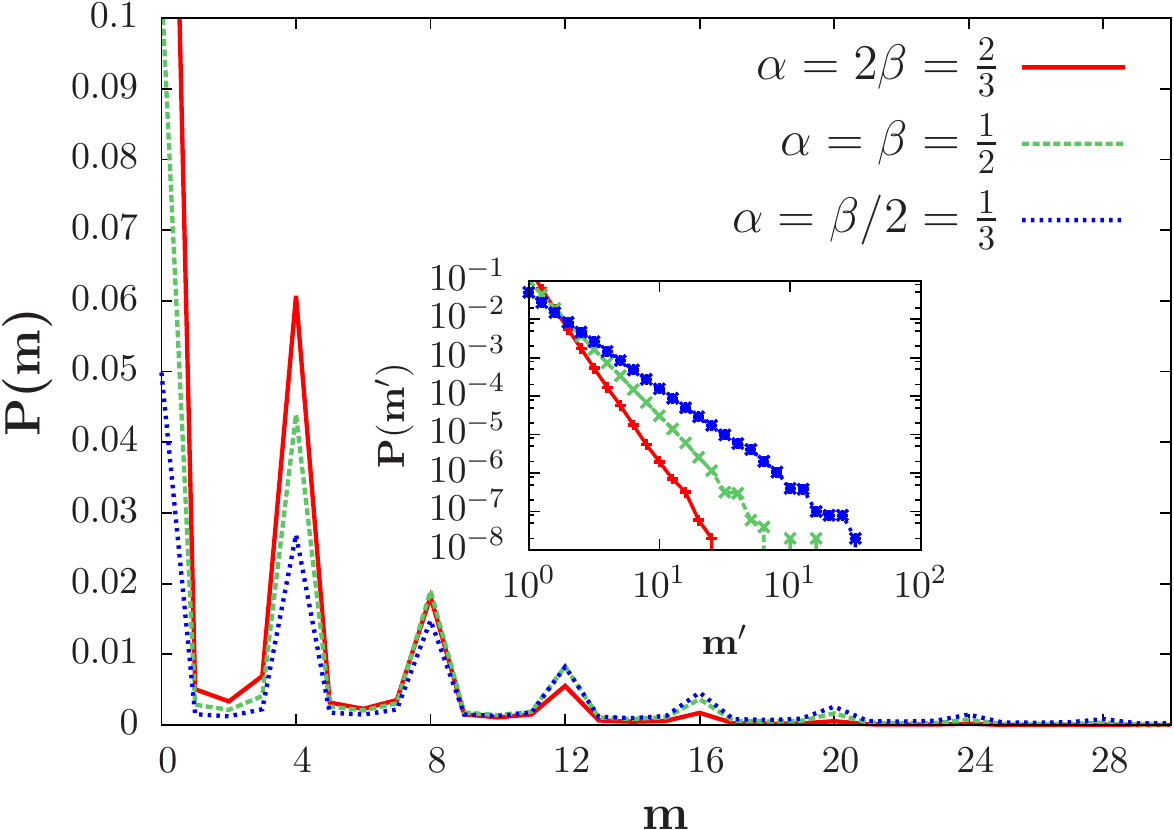}
    \caption{}
 \end{subfigure}
 \caption{(a) Gap distribution for different rates of attachment/detachment with evaporation and deposition. The gaps peak at multiples of $4$ for $P_{on}=0.8$ which corresponds to low load. For high-load conditions the curve decays smoothly. Inset shows exponentially decaying amplitudes of the peaks. The chosen probabilities of influx and escape are $\alpha=\beta=0.5$. (b) Gap distribution for different rates of influx and escape remain unaffected with the addition of deposition and evaporation. Chosen $P_{on}=S_{off}=0.8,P_{off}=S_{on}=0.2$.}
 \label{Figure:9}
\end{figure}

Comparing Fig.~\ref{Figure:8} to Fig.~\ref{Figure:6}, we notice that the damped oscillatory behavior at the entrance boundary remains unaffected by the addition of the Langmuir kinetics. From Fig.~\ref{Figure:8}(a), it is important to note that the data corresponding to high load ($P_{on}=0.2$) without Langmuir dynamics behaves almost like TASEP and hence the system is in the maximal current phase saturating at half density. Interestingly, adding Langmuir dynamics shifts the phase curve towards low density and it saturates at a much lower value of $\sim 0.05$. For low load conditions, $P_{on}=0.8$, the effect of deposition of particles dominate near the rear boundary of the lattice, leading to a positive slope in the density profile. This slight positive slope in the bulk density is consistently observed for all the three cases of influx and escape rates, see Fig.~\ref{Figure:8}(b). For the case with $\alpha=2\beta=2/3$, we observe that the density forms a cusp halfway through the lattice. We believe that this is an artifact of the competition between Langmuir dynamics and influx/escape dynamics. Even without the Langmuir dynamics, for lower escape rates $\beta$, the particle spends longer time on the lattice and hence they accumulate near the right boundary as seen in Fig~\ref{Figure:6}(b). Further, we hypothesize that adding evaporation balances this accumulation on the right boundary leading to the cusp formation. However, a detailed analytical study is required to understand this behavior which is beyond the scope of the current work. Analogous to the results in the previous section without Langmuir kinetics, the gap distribution loses its multiple peak structure for high load, $P_{on}=0.5$ and $0.2$ as a result of both higher load forcing the particles to take smaller steps and random deposition of particles destroying the gaps of four as can be seen from Fig.~\ref{Figure:9}(b).

\section{Conclusion and Outlook}
\label{Conc}
Our stochastic modeling of dynein motor transport on a lattice with the incorporation of varying hop sizes manifested many exciting results which include the oscillatory behavior of the \enquote{gap distribution} as well as the average density for the open boundary conditions, and superdiffusive growth of the fluctuations. Before probing into the combined effect of obstruction due to the presence of other motors and the impact of loads, we verified the experimental data for a single dynein motor given in the paper by Rai et al.~\cite{rai2013molecular} in our model. We found that without the traffic obstruction, a single dynein motor takes maximum jumps of step size $=4$. We also verified that the motor increasingly takes smaller steps and the velocity exponentially decreases with increasing load. Moreover, including the crowded track along with the load in our model, we observe the distribution of step sizes. For low load attachment case, the effect of the hindrance due to the presence of other motors is not that prominent, and motors still on an average prefer to jump four steps. Further, the stationary state current shows a breakdown of particle-hole symmetry and in the system exhibits free flow motion for densities smaller than $\approx 0.33$, depending on the choice of attachment and detachment probabilities. The intuitive features of the gap distribution allowed us to obtain a better insight into the collective behavior and self-organizing nature of the dyneins for both open and periodic boundary conditions. The gap distribution peaks at gap sizes, which are multiples of the maximum step size, signifying that the motors organize themselves so they can jump by the maximum step size possible. We also carried out a mean-field calculation for the gap distribution. For the more realistic scenario of open boundary conditions, our simulations capture a damped oscillating behavior of the density profile. The decay length highly depends on the attachment and detachment probabilities, as well as the influx and escape rates of the motors. Similar oscillations, albeit undamped unlike in our model, have been observed in earlier studies of dynein motors~\cite{Kunwar:2006} using cellular automata; here, we would like to emphasize the fact that different modeling approaches lead to very different results. The exact cause of this difference is an open question. For the simulations with Langmuir kinetics, the damped oscillations in the density profile and the discrete peaks in the gap distribution remained intact for low-load conditions. However, for high load, the distributions became smoother and imitated the results of TASEP. For the mean square displacement, we observe two regimes, an early time of fast growth followed by diffusive growth as the system approaches the stationary state. In the initial transient period, open and periodic boundary conditions show ballistic and superdiffusive growth with exponents $2$ and $1.3$, respectively. In the future, we would like to expand our minimalistic model by adding complexities that depict more experimentally relevant and realistic scenarios. A few possible adaptations can be limiting the amount of available ATP, probing the dynamics of walking on multiple protofilaments, etc. It would also be interesting to test these results in an in-vitro experimental set-up of an assembly of dynein motors. Although one of the major challenges for such an experimental study is controlling the number of motors and maintaining the composition of the motor assemblies, experiments using biosynthetic methods on DNA scaffolds ~\cite{rogers2009negative,furuta2013measuring,derr2012tug} have been conducted to investigate collective transport of motors. Our theoretical results illustrate how a group of dynein motors might be distributed on a single track depending on the load they carry, what would be the optimal density for them to most efficiently transport cargo and avoid over-crowding and thus, provide a significant point of reference for experiments on the collective motion of dyneins. 
\appendix
\numberwithin{equation}{section}
\makeatletter 
\section{Steady state gap distribution for $m_s=2$}
\label{appA}
We analytically solve for the gap distribution P(m) for the case when a motor can hop to maximally 2 empty sites in front of it. The rate of hopping $n$ steps with $m$ empty spaces in front of the motor is, 
{\color{red}\begin{equation}
    u_n(m)= [v_1\delta_{m,1}+v_2\theta(m-2)]\delta_{n,1} +  v_3\delta_{n,2}\theta(m-2)~.
\end{equation}}

 The master equations for the time dependent gap distribution $P(m,t)$ are,
\begin{eqnarray}
    \frac{dP(0,t)}{dt}&=& -P(0,t) \Big[v_1 P(1,t)+(v_2+v_3)\sum_{m=2}^\infty P(m,t)\Big] \no \\ && + v_1P(1,t)+ v_3P(2,t)~, \\
    \frac{dP(1,t)}{dt} &=& -P(1,t)\Big[ v_1(1+P(1,t)) + (v_2+ v_3)\sum_{m=2}^\infty P(m,t)\Big]+v_2 P(2,t) \no \\ && + v_3 P(3,t) + v_1 P(1,t)P(0,t) +v_2 \sum_{m=2}^\infty P(m,t) P(0,t)~, \\
    \frac{dP(m,t)}{dt} &=& -P(m,t)(v_2+v_3) -P(m,t)[v_1 P(1,t)+(v_2+v_3)\sum_{m=2}^\infty P(m,t)] \no \\ && +v_2 P(m+1,t) + v_3 P(m+2,t) + P(m-1,t)[v_1 P(1,t) \no \\ && +v_2\sum_{m=2}^\infty P(m,t)]+ v_3 P(m-2)\sum_{m=2}^\infty P(m,t)~.
\end{eqnarray}
Since the total probability is conserved, $\displaystyle \sum_{m=0}^\infty P(m,t)=1$, defining $\displaystyle  \sum_{m=k}^\infty P(m,t)=S_k$ and using the conservation law $ P(0,t) + \sum_{m=1}^\infty P(m,t)=1$, we have
\begin{equation}
    P(0,t)=1-S_1~. 
\end{equation}
Similarly, $P(1,t)$ can be represented as $S_1-S_2$. Using these substitutions, we arrive at the simplified form of the evolution equation,

\begin{eqnarray}
 \frac{dP(0,t)}{dt}&=& -P(0,t) \Big[v_1 P(1,t)+(v_2+v_3)S_2\Big]+ v_1P(1,t) +v_3P(2,t) ~,\\
    \frac{dP(1,t)}{dt} &=& -P(1,t)\Big[ v_1(1+S_1-S_2) + (v_2+ v_3)S_2\Big] \no \\
    && +v_2 P(2,t) + v_3 P(3,t) + \Big[v_1(S_1-S_2)+v_2S_2\Big]P(0,t)~, \\
    \frac{dP(m,t)}{dt} &=& -P(m,t)\Big[(v_2+v_3)(1+S_2)+v_1(S_1-S_2)\Big]+ v_2P(m+1,t)  \no \\
 &&+ v_3P(m+2,t) + v_3S_2P(m-2,t) \no\\ &&+ P(m-1)[v_1(S_1-S_2)+v_2S_2]~ \qquad m\ge2.
\end{eqnarray}

 Defining the generating function, $\displaystyle G=\sum_{m=1}^\infty z^m P(m,t)$, we can write the evolution equation as,
 \begin{eqnarray}
  \frac{dG}{dt}&=&\frac{d}{dt}\sum_{m=1}^\infty z^m P(m,t) \nonumber \\
    &=&\Big[[(1+S_2)(v_2+v_3)+v_1(S_1-S_2)]+w_2z^{-1}+w_3z^{-2}\no \\
    &&+[v_1(S_1-S_2)+w_2S_2]z+w_2S_2z^2\Big]G
+z(v_2+v_3-v_1)P(1,t)\no\\&&+v_2 P(1,t)-w_3z^{-1}P(1,t)-v_3 P(2,t)\no\\&&+[v_1(S_1-S_2)+w_2S_2]zP(0,t)~.
\label{GF}
 \end{eqnarray}  
 In the stationary state, the probabilities are time independent. Hence, setting $d P(0,t)/dt$ equal to zero, one can obtain the following boundary condition,
\begin{equation}
    v_3 P(2)=-v_1 P(1)+P(m)[v_1 P(1)+(v_2+v_3)S_2]~.
\label{P2}
\end{equation}
 To solve for stationary state distribution, we set $dG/dt=0$, and after plugging the expression Eq.~\ref{P2} of $P(2)$ in Eq.~\ref{GF} which gives us the following polynomial equation,
\begin{align}
 &\Big[-z^{4} v_3 S_2 -z^{3}[v_1(S_1-S_2)+v_2S_2]+z^2[(1+S_2)(v_2+v_3)+v_1(S_1-S_2)]\no\\&-v_2 z-v_3\Big]G=z^4v_3S_2(1-S_1)+z^3\Big[(v_2+v_3-v_1S_1)(S_1-S_2)v_2S_2(1-S_2)\Big]\no\\&-z^2\Big[[v_1(S_1-S_2)+(v_2+v_3)S_2](1-S_1)+(v_1-v_2)(S_1-S_2)-v_3(S_1-S_2)z~.
\end{align}
\begin{figure}
\centering
    \includegraphics[width=0.7\columnwidth]{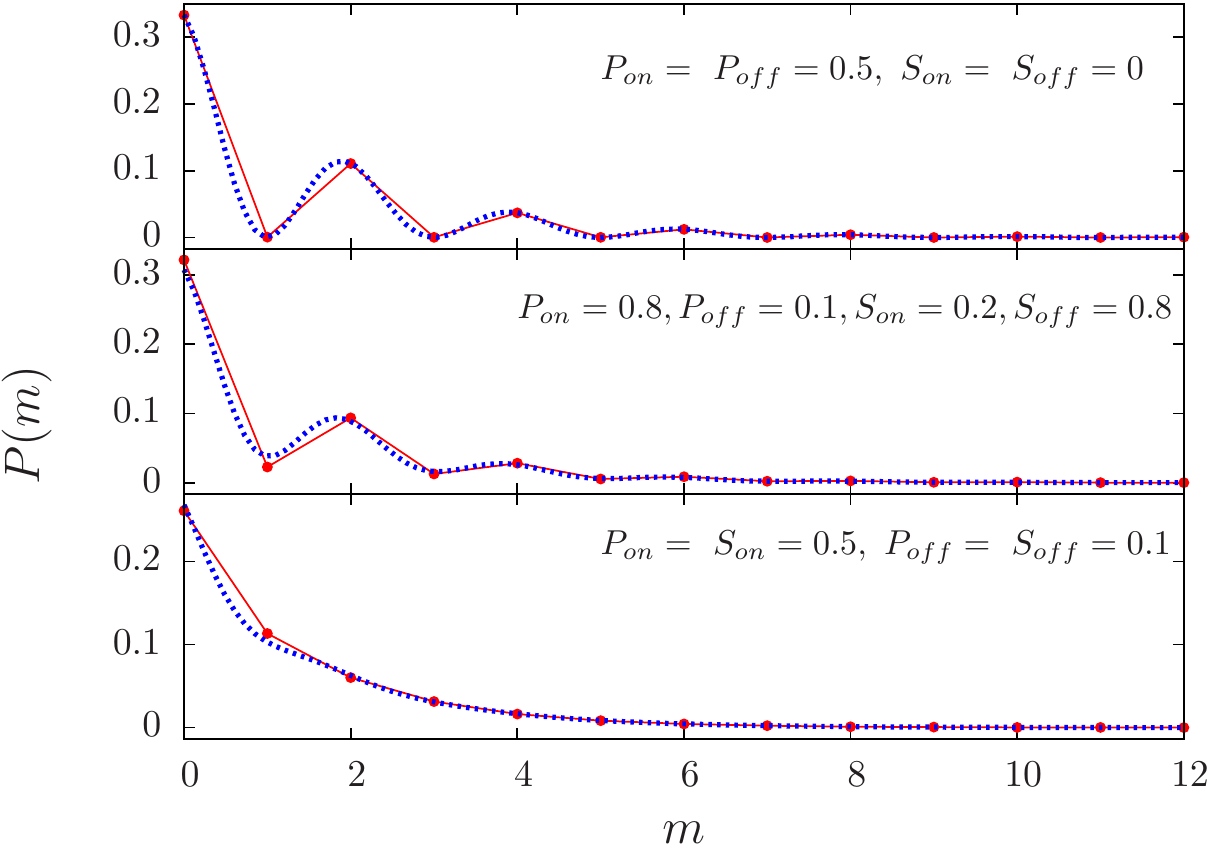}
    \caption{The gap size distribution for $m_s=2$ for various probabilities of attachment and detachment. The analytical result (\ref{cubic_eq}) represented by dashed lines matches very well with our simulation results (line-point),}
\label{Plotm_s2}
\end{figure}
Simplifying the above equation by taking out the common factor of $(z-1)$, we get
\begin{align}
&\Big[-z^3 v_3S_2-z^2[(v_2+v_3)S_2+v_1(S_1-S_2)]+z(v_2+v_3)+v_3\Big]G=\no\\&z^3 v_3S_2(1-S_1)+z^2[(v_2+v_3)S_1(1-S_2)-v_1S_1(S_1-S_2)]+zv_3(S_1-S_2)~.
\label{SolGF}
\end{align}
Further, we expand $G$, equate all coefficients for $z^m$ from both sides of (\ref{SolGF}), and obtain the recursion relation for $P(m)$ which is also calculated in \cite{Shrivastav:2010}. In case of $m_s=2$, the recurrence relation for $P(m)$ is given as
\begin{align}
    P(m)v_3 + P(m-1)(v_2 +v_3)- P(m-2)(S_1v_1 - S_2(v_2+ v_3-v_1)-P(m-3)S_2v_3=0~. 
\label{poly}
\end{align}

Assuming a polynomial solution, $P(m)~x^{m}$ and using in (\ref{poly}), we get a third order polynomial in $x$,
\begin{equation}
 v_3 x^3+(v_2+v_3)x^2-(S_2(v_2+v_3-v_1)+S_1 v_1)x-S_2v_3=0~.
\label{cubic_eq}
\end{equation} 
 Solution of above equation has three roots $x_1,x_2,x_3$ and hence $P(m)=c_1x_1^m+c_2x_2^m+c_3x_3^m$. Using conservation laws Eqs.~\ref{Crule1} and \ref{Crule2}, we consider that the coefficients $c_3=0$ and $c_1,c_2$ values depend on the rates. We solve two consecutive equations for specifying the values of the transition rates, $v_1,v_2,v_3$ and the density the $\rho$ and solved for all the unknowns. We have shown the comparison of the solution with Monte-Carlo simulation of $N$-motor system in Fig.~\ref{Plotm_s2}.
 $v_i$ depends on the attachment and detachment rates as,
 \begin{eqnarray}
     v_2=P_{on}(1-P_{off})S_{on}(1-S_{off}) ~, \no \\ 
     v_3 = P_{on}(1-P_{off})[(1 - S_{on})+ S_{on}S_{off}]~, \no \\
     v_1 = v_3 \rho + v_2 ~.  
     \label{rate_rel}
 \end{eqnarray}

\end{document}